\documentstyle[twocolumn,aps,prb]{revtex}
\begin{document}

\draft
\title
{\bf Fano interference effect on the transition spectrum of single
electron transistors}

\author{ David M.-T. Kuo}
\address{Department of Electrical Engineering, National Central
University, Chung-Li, Taiwan, 320, Republic of China}

%\end{center}
\date{\today}
\maketitle

\begin{abstract}
We theoretically study the intraband transition spectrum of single
electron transistors (SETs) composed of individual self-assembled
quantum dots. The polarization of SETs is obtained by using the
nonequilibrium Green's function technique and the Anderson model
with three energy levels. Owing to nonradiative coupling between
two excited states through the continuum of electrodes, the Fano
interference effect significantly influences the peak position and
intensity of infrared wavelength single-photon spectrum.
\end{abstract}

%\newpage
\section{Introduction}
The quantum dot (QD) system has many potential applications in
electronic devices, including quantum dot lasers$^{1}$ and
infrared detectors$^{2}$. Recently, the spontaneous emission
spectrum of QDs embedded in a semiconductor p-n junction has been
proposed as a generator of single photons, which is important in
the application of quantum cryptography$^{3}$. The antibunching
feature of a single photon source was demonstrated by optical and
electrical pumping, respectively$^{4-5}$. Thus far, the
implementation of a single photon source has used short
wavelengths (near $1.3\mu m$). For such a short wavelength photon
can be generated by electron-hole recombination in the excitations
including exciton, negative trion, positive trion and biexciton
states$^{6}$.

In some applications of quantum communications, a long-wavelength
single photon near a $10 \mu m$  infrared wavelength is useful
owing to its advantage of high transmission in the atmosphere.
Note that a long wavelength infrared single photon passes through
the atmosphere in the $3-5 ~\mu m$ and $8-14 ~\mu m$ bands$^{7}$,
the atmospheric transmission is near $0.98$ in the infrared range
from $9.7 ~\mu m$ to $10.2~ \mu m$. To generate a long-wavelength
single photon, we can employ electron intraband transitions. One
way of realizing such photon source is to use single electron
transistors (SETs), which consists of two leads and one QD. Owing
to Coulomb blockade effect and Pauli principle, the SETs can be
utilized not only as a single photon emitter, but also a single
photon detector.

To date, many nanostructure materials such as silicon (Si) and
germanium (Ge) QDs can be miniaturized by using advanced
fabrication technology. Furthermore, recently, the Ge/Si SETs have
been demonstrated at room temperature$^{8,9}$.Due to Si and Ge are
multi-valley semiconductors, we propose to embed a single
self-assembled InAs QD into GaAs slab (with width W) which is then
placed in contact with electrodes to form n-i-n structure. In the
fabrication of optoelectronic devices such as single-photon
generators and single-photon detectors, material for single valley
semiconductors are preferred. Even though it is difficult to align
a single dot with the source and drain electrodes, this difficulty
may be solved by the selective formation method$^{10,11}$, which
can improve position control and minimize the size fluctuation of
self-assembly quantum dot (SAQD). To produce a single photon near
a $10 ~\mu m$ infrared wavelength, first we need to know the
energy levels of a single InAs QD, which depend on the shape of
QDs. Here the pyramid-shaped InAs QDs with the base length b and
height h will be considered since this shape is commonly observed
in many experiments.

Within the effective mass model$^{6}$, the lowest three energy
levels for electrons as a function of base length are shown in
Fig. 1, where the ratio of $h/b=1/4$ is considered. The first
excited state $E_p$ is a fourfold degenerate state including spin,
whereas this degenerate state will be lifted as $E_2$ and $E_3$ if
the x-y axis symmetry of QD is destroyed. Fano interference could
be occurred due to coupling of these states to the same
reservoir$^{12}$. The main purpose of this article is to
theoretically investigate Fano interference effect on the
transition spectrum of single photons near a $10 ~\mu m $ (or $
124~ meV)$ infrared wavelength.

\section{ Hamiltonian of system}
We start with the Anderson model with three energy levels to
describe the studied system

\begin{eqnarray}
H_{new} & = & \sum_{{\bf k}, \sigma, \alpha} \epsilon_{{\bf k}}
c^{\dagger}_{{\bf k},\sigma,\alpha} c_{{\bf k}, \sigma,\alpha}
+\sum_{\sigma,\ell=1,2,3} E_{\ell} d^{\dagger}_{\ell,\sigma} d_{\ell,\sigma} \nonumber \\
 &   & +\sum_{{\bf k}, \sigma,\alpha,\ell} (V_{{\bf k}, \alpha,\ell}
c^{\dagger}_{{\bf k},\sigma,\alpha} d_{\ell,\sigma}
+V^{\dagger}_{{\bf k},\alpha,\ell} d^{\dagger}_{\ell,\sigma}
c_{{\bf k}, \sigma,\alpha} )\\ \nonumber &+
&\sum_{\ell=2,3}(\lambda_{\ell,1}e^{i\omega t}
d^{\dagger}_{\ell,\sigma} d_{1,\sigma} +
\lambda^{\dagger}_{\ell,1} e^{-i\omega t}
d^{\dagger}_{1,\sigma}d_{\ell,\sigma}),
\end{eqnarray}
where $ c^{\dagger}_{{\bf k},\sigma,\alpha}$ and
$d^{\dagger}_{\ell,\sigma}$ are the electron creation operators in
the electrodes and QDs. The first term describes the left and
right electrodes via index $\alpha=L, R$. The second term
describes the electronic states of QDs. We consider the situation
where the QD contains three bound levels ($\ell=1,2,3$). The third
and fourth terms describe the coupling between the QD states and
the two electrodes. The last two terms describe the interaction of
the QD electrons with electromagnetic field. Here $\lambda =
-\mu_r {\cal E} $ is the Rabi frequency, where $\mu_r=|\langle
\ell|{\bf r}|1\rangle |$ is the matrix element for the intraband
transition and ${\cal E}=(\omega/\epsilon_0 V)^{1/2}$ is electric
field per photon. $V$ and $\epsilon_0$, respectively, are the
volume and static dielectric constant of the system. We have used
the units such that $\hbar=c=1$. This convention is used
throughout this paper. We introduce a unitary transformation
$S(t)= e^{i(\omega t/2)[d^{\dagger}_{1,\sigma}
d_{1,\sigma}-d^{\dagger}_{2,\sigma}
d_{2,\sigma}-d^{\dagger}_{3,\sigma} d_{3,\sigma}]}$ and define the
transformed Hamiltonian by $H_{new} =
S^{\dagger}(t)H(t)S(t)-S^{\dagger}(t)i\partial/\partial t S(t)$.
The new Hamiltonian takes the form

\begin{eqnarray}
H_{new} & = & \sum_{{\bf k}, \sigma, \alpha} \epsilon_{{\bf k}}
c^{\dagger}_{{\bf k},\sigma,\alpha} c_{{\bf k}, \sigma,\alpha}
+\sum_{\sigma,\ell=1,2,3} \epsilon_{\ell} d^{\dagger}_{\ell,\sigma} d_{\ell,\sigma} \nonumber \\
 &   & +\sum_{{\bf k}, \sigma,\alpha,\ell} (V_{{\bf k},
 \alpha,\ell}(t)
c^{\dagger}_{{\bf k},\sigma,\alpha} d_{\ell,\sigma}
+V^{\dagger}_{{\bf k},\alpha,\ell}(t) d^{\dagger}_{\ell,\sigma}
c_{{\bf k}, \sigma,\alpha} )\\ \nonumber &+
&\sum_{\ell=2,3}(\lambda_{\ell,1} d^{\dagger}_{\ell,\sigma}
d_{1,\sigma} + \lambda^{\dagger}_{\ell,1}
d^{\dagger}_{1,\sigma}d_{\ell,\sigma}),
\end{eqnarray}
where the renormalized energy levels of QD are $\epsilon_{1}=
E_{1}+ \omega/2$ and $\epsilon_{\ell=2,3}= E_{\ell}-\omega/2 $. We
see that the time-dependent phase in the inter-level Hamiltonian
vanishes. However, the hopping terms are time dependent, $V_{{\bf
k},1}(t)= V_{{\bf k},1} exp^{i\omega/2 t}$ and $V_{{\bf
k},j=2,3}(t)= V_{{\bf k},j} exp^{-i\omega/2 t}$, in which the
energy and time dependence of the coupling are factorized. This
factorization leads to time-independent tunneling rates. Even
though the effect of electron correlation is significant in small
semiconductor QDs, we can ignore the particle Coulomb interaction
if the applied voltage is not sufficient to overcome the charging
energies of QDs$^{6}$. Such condition is required throughout this
article. According to Eq. (2), two excited states are indirectly
coupled via the continuum of electrodes. This coupling can be
enhanced by including electron-phonon interaction. In this study
we phenomenologically include electron-phonon interaction via the
complex self-energy of $\kappa_{2,3} = \Delta_{d}+
i\sqrt{\Gamma_{d2} \Gamma_{d3}}$, which was considered to study
double quantum wells in Ref. [13]. $\Gamma_{d2}$ and $\Gamma_{3d}$
denote the decay rates of $E_2$ and $E_3$. $\Delta_d$ is the
coupling strength between $E_2$ and $E_3$.  Note that although the
electron-phonon scattering rate can be small or vanishing in QDs
due to the phonon bottleneck effect (inability to satisfy the
energy conservation between discrete QD levels), it can still give
rise to significant contribution to the phonon-assisted tunneling,
since the energy conservation can be met by electron jumping
between the QD and the leads.

The electrically driven transition spectrum of individual SETs can
be calculated by using the Keldysh-Green's function
method$^{14,15}$.The lesser Green's function
$G^{+,-}_{1,2}(t_1,t_2)=
-i<d^{\dagger}_{2,\sigma}(t_2)d_{1,\sigma}(t_1)>$, which describes
the correlation of electron in the energy level $\epsilon_1 $ and
$\epsilon_2 $ at time $t_1$ and $t_2$, is introduced to calculate
the polarization. Using the method of equation of motion, the
stationary solution is given by
\begin{small}
\begin{eqnarray}
& &
(\epsilon_2-\epsilon_{1}+i(\Gamma_{1}+\Gamma_2))G^{+,-}_{1,2}(\epsilon)
\\ \nonumber &=&-i\lambda_{1,2}
[G^{+,-}_{1,1}(\epsilon)-G^{+,-}_{2,2}(\epsilon)]
+ \Sigma^{<}_1 G^a_{1,2}(\epsilon+\frac{\omega}{2})\\
\nonumber &-& \Sigma^{<}_{2}
G^r_{1,2}(\epsilon-\frac{\omega}{2})-\kappa_{2,3}
G^{+,-}_{1,3}(\epsilon)
\end{eqnarray}
\end{small}
In quasi-equilibrium the intra-level lesser self energies of
Eq.(3) are $\Sigma^{<}_i=\Gamma_i
(f_{L}(\epsilon)+f_{R}(\epsilon))$, where $f_{L}(\epsilon)$ and
$f_{R}(\epsilon)$ are the Fermi distribution function of the left
and right electrodes. The chemical potential difference between
these two leads is related to the applied bias $\mu_L-\mu_R=2~e
V_a$. $\Gamma_{\ell,\alpha=L,R}= \sum_{\bf k} |V_{{\bf
k},\ell,\alpha}|^2 \delta(\epsilon-\epsilon_{\bf k})$ denote the
tunneling rates from the QD to the left and right electrodes,
respectively. For simplicity, we consider
$\Gamma_{i,L}=\Gamma_{i,R}=\Gamma_{i}$. It is very difficult to
fully include the tunneling rate as a function of energy and
momentum, and we assume that these tunneling rates are energy and
bias independent even though $\Gamma_1 (\Gamma_2)$ can be
calculated with a reliable method$^{16}$.

The lesser Green's functions $G^{+,-}_{1,1}(\epsilon)$ and
$G^{+,-}_{2,2}(\epsilon)$ of Eq. (3) denote the electron
occupation numbers, which are determined by the spectrum functions
$ G^{+,-}_{1,1}(\epsilon) =
f^{<}_1(\epsilon)A_{1}(\epsilon+\frac{\omega}{2})$ and
$G^{+,-}_{2,2}(\epsilon) = f^{<}_2(\epsilon)
A_{2}(\epsilon-\frac{\omega}{2})$. The quasi-equilibrium
distribution function of QD is expressed by
$f^{<}_i(\epsilon)=(\Gamma_{iL}f_{L}(\epsilon)+\Gamma_{iR}f_{R}(\epsilon))/(\Gamma_{iL}+\Gamma_{iR})$
. Owing to the weak electron-photon coupling, we make the
approximations of the spectrum functions
$A_{1(2)}(\epsilon)\approx \pm 2 ImG^{r(a)}_{1(2),0}(\epsilon)$,
which are the imaginary part of Green's functions, where the
retarded and advanced Green functions are
$G^{r(a)}_{1,0}(\epsilon)=1/(\epsilon-\epsilon_1 \pm i
\Gamma_{1})$, and $G^{r(a)}_{2,0}(\epsilon)=
1/(\epsilon-\epsilon_2 \pm i\Gamma_{2})$. Such approximation
implies that the electron occupation numbers are mainly
contributed from the tunneling process. In Eq.(3) the interlevel
Green's function is given by $G^{a(r)}_{1,2}(\epsilon) =
\lambda_{1,2} G^{a(r)}_{1,0}(\epsilon) G^{a(r)}_{2,0}(\epsilon)$.
It is worthy noting that we ignored the term of $\lambda_{1,3}
G^{+,-}_{3,2}(\epsilon)$ in Eq. (3), where
$G^{+,-}_{3,2}=\kappa_{2,3}[f^{<}_2(\epsilon)A_2(\epsilon)G^r_{3,0}(\epsilon)
+f^{<}_3(\epsilon)A_3(\epsilon) G^a_2(\epsilon)]$, which can be
readily proved as a negligible term.

Defining frequency detuning as $\Delta
\omega_{12}=\epsilon_2-\epsilon_1 + i
(\Gamma_1+\Gamma_2)=1/G^r_{1,0}(\epsilon)-1/G^a_{2,0}(\epsilon)$
and using
$2\Gamma_{i}=-i(\frac{1}{G^r_{i,0}(\epsilon)}-\frac{1}{G^{a}_{i,0}(\epsilon)})$
, we rewrite Eq. (3) as
\begin{small}
\begin{eqnarray}
G^{+,-}_{1,2}(\epsilon) & =& \lambda_{1,2} (f^{<}_{2}(\epsilon)
A_{2,0}(\epsilon-\frac{\omega}{2})G^{r}_{1,0}(\epsilon-\frac{\omega}{2})
\\ \nonumber &+&f^{<}_{1}(\epsilon) A_{1,0}(\epsilon+\frac{\omega}{2})
G^{a}_{2,0}(\epsilon+\frac{\omega}{2}))-\frac{\kappa_{2,3}}{\Delta \omega_{12}} G^{+,-}_{1,3}(\epsilon),\\
\nonumber & = & P_1(\epsilon)-\frac{\kappa_{2,3}}{\Delta
\omega_{12}} G^{+,-}_{1,3}(\epsilon).
\end{eqnarray}
\end{small}
In the absence of $\kappa_{2,3}$, $G^{+,-}_{1,2}(\epsilon)$
describes the transition spectrum between the ground state $E_1$
and the first excited state $E_2$, which is viewed as the first
line shape. The self energy $\kappa_{2,3}$ leads
$G^{+,-}_{1,2}(\epsilon)$ to couple with  Green's functions
$G^{+,-}_{1,3}(\epsilon)$, which is expressed by
\begin{small}
\begin{eqnarray}
G^{+,-}_{1,3}(\epsilon) & =& \lambda_{1,3} (f^{<}_{3}(\epsilon)
A_{3,0}(\epsilon-\frac{\omega}{2})G^{r}_{1,0}(\epsilon-\frac{\omega}{2})
,
\\ \nonumber & +&  f^{<}_{1}(\epsilon) A_{1,0}(\epsilon+\frac{\omega}{2})
G^{a}_{3,0}(\epsilon+\frac{\omega}{2}))-\frac{\kappa^*_{2,3}}{\Delta
\omega_{13}} G^{+,-}_{1,2}(\epsilon)\\ \nonumber &= &
P_2(\epsilon)- \frac{\kappa^*_{2,3}}{\Delta \omega_{13}}
G^{+,-}_{1,2}(\epsilon),
\end{eqnarray}
\end{small}
where $\Delta \omega_{13} = \epsilon_3 - \epsilon_1 + i (\Gamma_1
+\Gamma_3)=1/G^r_{1,0}(\epsilon)-1/G^a_{3,0}(\epsilon)$. $P_2$
describes the second line shape. This second line shape will
interfere with the first line shape due to $\kappa_{2,3}$. Solving
Eqs.(4) and (5), we obtain the polarization consisted of two
components $X_{12}(\omega)=\int d\epsilon/(2\pi) \lambda^{*}_{1,2}
G^{+,-}_{1,2}(\epsilon) $ and $X_{13}(\omega)= \int
d\epsilon/(2\pi) \lambda^{*}_{1,3} G^{+,-}_{1,3}(\epsilon)$:
\begin{equation}
X_{12}(\omega) = \int \frac{d\epsilon}{2\pi} \frac{
\lambda^*_{1,2}\Delta
\omega_{1,3}(-P_2(\epsilon)\kappa_{2,3}+P_1(\epsilon) \Delta
\omega_{1,2})} {\Delta \omega_{1,3}\Delta \omega_{1,2} -
|\kappa_{2,3}|^2},
\end{equation}
and
\begin{equation}
X_{13}(\omega)= \int \frac{d\epsilon}{2\pi} \frac{\lambda^*_{1,3}
\Delta \omega_{1,2}(-P_1(\epsilon)\kappa^{*}_{2,3}+ P_2(\epsilon)
\Delta \omega_{1,3})}{\Delta \omega_{1,3} \Delta \omega_{1,2} -
|\kappa_{2,3}|^2}.
\end{equation}

Eqs. (6) and (7) are the central results of this article. Based on
Eqs. (6) and (7), we will investigate the Fano interference effect
arising from $\kappa_{2,3}$ on the absorption and emission
spectrum, respectively.

\section{Results and discussions}
To generate a single photon at the $10 ~\mu m$ wavelength, the QD
with base length $b \approx 14~nm$ is sufficient to provide such
photons ($\omega=E_2-E_1 \approx 124~meV$). To simulate the
system, we also choose the Fermi energy $E_F=50~ meV$ which is
$30~ meV $ below $E_1$ at zero bias. In addition, we assume
$E_3-E_2=6~meV$. Therefore, the levels of QD are empty at zero
bias. When the chemical potential of left electrode $\mu_L$
(source) sweeps through the ground state and far away from the
excited states (the energy level separation $\Delta E=E_2-E_1 \gg
k_B T$), Eqs. (6) and (7) can be rewritten as the following
expression
\begin{eqnarray}
X_a& = & {\cal P}_1(\omega) \int \frac{d \epsilon}{2\pi}
f^{<}_{1}(\epsilon) G^a_{2,0}(\epsilon+\frac{\omega}{2})
A_{1,0}(\epsilon+\frac{\omega}{2}) \\
\nonumber &+& {\cal P}_2(\omega)\int \frac{d \epsilon}{2\pi}
f^{<}_{1}(\epsilon)G^a_{3,0}(\epsilon+\frac{\omega}{2})
A_{1,0}(\epsilon+\frac{\omega}{2}),
\end{eqnarray}
where ${\cal P}_1=(|\lambda_{1,2}|^2\Delta \omega_{1,2} \Delta
\omega_{1,3} - \lambda_{1,2} \lambda^{*}_{1,3} \kappa_{2,3} \Delta
\omega_{1,2})/D$, ${\cal P}_2= (|\lambda_{1,3}|^2 \Delta
\omega_{1,2} \Delta \omega_{1,3} - \lambda^{*}_{2,1}\lambda_{1,3}
\kappa_{2,3} \Delta \omega_{1,3})/D$ and $D=\Delta \omega_{1,3}
\Delta \omega_{1,2} -|\kappa_{2,3}|^2$. Owing to very small
tunneling rate of $\Gamma_1$ for deep energy level, the expression
of Eq. (8) at zero temperature has a simple form

\begin{equation}
X_a =  \frac{1}{2}\frac{|\lambda_{1,2}|^2  \Delta \omega_{1,3} +
|\lambda_{1,3}|^2 \Delta \omega_{1,2}- 2\lambda_{1,3} \kappa_{2,3}
\lambda_{2,1}}{\Delta \omega_{1,3} \Delta \omega_{1,2}
-|\kappa_{2,3}|^2}.
\end{equation}

The expression of Eq. (9) was also obtained for atomic systems to
study electromagnetically induced transparency (EIT) $^{17,18}$.
Fig. 2 shows the imaginary part of $X_a(\omega)$ for
$\kappa_{2,3}=i~meV$ and $\Gamma_2=\Gamma_3=1~ meV$. The solid
line and dashed line denote, respectively, with and without
$\kappa_{2,3}$. We see that the Fano interference effect arising
from $\kappa_{2,3}$ significantly influences the absorption
spectrum. It is destructive between two peaks, but constructive in
the wings of the two absorption lines. This EIT effect can be used
to efficiently modulate the light group velocity (or stop
light)$^{12,19}$ as a result of very small broadening of energy
levels for QDs. To date, this quantum interference has not been
reported in isolated QD system, whereas it was reported in the
quantum well system$^{20}$. Although Eq. (8) can provide further
information about the absorption spectrum for the different
applied voltages and temperatures, we will discuss these effects
on the emission spectrum.

When the left lead supplies electrons into the excited states,
light emission process occurs. Emission spectrum intensity is
given by
\begin{eqnarray}
X_e(\omega) & = & {\cal P}_1(\omega) \int \frac{d \epsilon}{2\pi}
f^{<}_2(\epsilon) A_{2,0} (\epsilon-\frac{\omega}{2}) G^r_{1,0}(\epsilon-\frac{\omega}{2})\\
\nonumber & + &{\cal P}_2(\omega) \int \frac{d \epsilon}{2\pi}
f^{<}_3(\epsilon) A_{3,0} (\epsilon-\frac{\omega}{2})
G^r_{1,0}(\epsilon-\frac{\omega}{2}).
\end{eqnarray}
 We show the imaginary part of $X_e(\omega)$ for the
applied voltage $V_a=165 ~mV$ and temperature $k_B T=1~meV$ in
Fig. 3: the solid line and dashed line denote, respectively,
$k_{2,3}=(-0.5+i)~meV$ and $\kappa_{2,3}=0$. Comparing to
absorption spectrum shown in Fig. 2, we see two asymmetric lines
resulting from the real part of $\kappa_{2,3}$, which was ignored
in the absorption spectrum. In addition to the destructive
interference between two peaks,the constructive interference is
also observed in the wings of these two peaks. This feature is the
same as that of absorption spectrum. However, note that the
intensity of $L_1$ is enhanced and slightly shifted from
$\omega=124~meV$. The intensity enhancement of $L_1$ implies that
the number of single photons emitted from $L_1$ is enhanced.
Therefore, we can take this advantage to the application of
quantum cryptography.

It is significant to provide single-photon sources at room
temperature for the development of quantum cryptography.
Therefore, we attempt to understand temperature effect on the
emission spectrum. We show the imaginary part of $ X_e(\omega)$
for different temperatures at applied voltage $V_a=165~mV$ in Fig.
4: solid line ($k_BT=1~meV$), dashed line ($k_BT=2~meV$), dotted
line($k_BT=3~meV$) and dash-dotted line ($k_BT=4~meV$). When the
applied voltage is $V_a=165~ mV$, the left electrode supplies
maximum electron number into two resonant levels $E_2$ and $E_3$
at zero temperature. Consequently, $L_1$ and $L_2$ display the
strongest intensities at the temperature of $k_BT=1~meV$. The
peaks $L_1$ and $L_2$ decline in heights, as the temperature
increases. Therefore, high temperature diminishes the efficiency
of photon emission. Besides, thermionic emission not included in
this study could also seriously destroy photon emission efficiency
at room temperature since the energy levels $E_2$ and $E_3$ are
not very deep (See Fig. 1).

From experimental point of view, the behavior of emission spectrum
versus applied voltage is also of interest. The imaginary part of
$X_e(\omega)$ for different applied voltages at temperature $k_B
T=1~meV$ is shown in Fig. 5: $V_a=159~mV$ (dash-dotted line),
$V_a=161~mV$ (dotted line), $V_a=163~mV$ (dashed line), and
$V_a=165~mV$ (solid line). When the applied voltage is $V_a=159
~mV$, electrons are injected into the resonant energy level $E_2$
to emit photons with $\omega \approx 124~meV$. In addition to
$L_1$ line shape, we also observe the emission line of $L_2$ with
frequency $\omega \approx 130~ meV$, which is mainly contributed
from $\kappa_{2,3}$.When the applied voltage is $V_a=161~mV$, the
intensity of $L_2$ is greatly enhanced as a result of that the
left electrode provides electrons into the resonant energy level
of $E_3$.The intensity of $L_2$ approaches saturation at $V_a=165~
mV$. From the results of Fig. 5, we have demonstrated SETs as
single-photon sources and double-photon sources. The latter may be
useful for the study of teleportation.$^{21}$

\section{Summary}
In this study we have proposed to utilize the intraband transition
of InAs/GaAs SETs to produce a single-photon source in the
infrared wavelength range for the application of wireless quantum
cryptography. The Anderson model with three energy levels is used
to simulate the studied system. It is found that quantum
interference effect leads two asymmetry line-shapes. It is
destructive between two peaks, but constructive in the wings of
two transition lines. The expression of Eqs. (5) and (6) can be
readily extended to the case of double quantum dot system or
quantum well system, where the bond and antibond states
correspond, respectively, to $E_2$ and $E_3$.

{\bf ACKNOWLEDGMENTS}

This work was supported by National Science Council of Republic of
China Contract No. NSC 94-2215-E-008-027.

%\section{Appendix A}

\mbox{}

%\newpage

{\bf Figure Captions}

Fig. 1: The lowest three energy levels of a quantum dot (QD) as
functions of the QD size b. $E_s$ and $E_p$ denote, respectively,
the ground state and the first excited state.

Fig. 2: Imaginary part of $X_a(\omega)$ as functions of detuning
frequency.

Fig. 3: Imaginary part of $X_e(\omega)$ as functions of detuning
frequency at applied voltage $V_a=165~ mV$ and temperature
$k_BT=1~meV$. Solid line ($\kappa_{2,3}=(-0.5+i)~ meV$) and dashed
line ($\kappa_{2,3}=0$).

Fig. 4: Imaginary part of $X_e(\omega)$ as functions of detuning
frequency for different temperatures at applied voltage $V_a=165~
mV$ and $\kappa_{2,3}=(-0.5+i)~ meV$.

Fig. 5: Imaginary part of $X_e(\omega)$ as functions of detuning
frequency for different applied voltages at temperature $k_B
T=1~meV$ and $\kappa_{2,3}=(-0.5+i)~ meV$.

%\end{small}

\end{document}